\documentstyle[11pt]{article}

\def\be{\begin{equation}}
\def\bea{\begin{eqnarray}}
\def\ee{\end{equation}}
\def\eea{\end{eqnarray}}
\def\ri{\rightarrow}
\def\ov{\overline}
\def\ra{\rangle}
\def\la{\langle}
\def\r{\right}
\def\l{\left}
\def\pa{\partial}

\def\a{\alpha}
\def\b{\beta}
\def\g{\gamma}
\def\d{\delta}
\def\eps{\epsilon}

\def\th{\theta}
\def\lam{\lambda}
\def\m{\mu}
\def\n{\nu}
\def\rh{\rho}
\def\s{\sigma}
\def\t{\tau}

\newcommand{\E}{{\rm e}}
\newcommand{\D}{{\rm d}}

\begin{document}
\title{Introduction to the statistical theory of Darwinian evolution\thanks
{Lectures given at the ICTP Summer College on frustrated systems, Trieste,
August 1997. Notes taken by Ugo Bastolla and Susanna Manrubia.}} 
\author{Luca Peliti\thanks{Associato INFN, Sezione di Napoli. 
E-mail: {\tt peliti@na.infn.it}} \\
Dipartimento di Scienze Fisiche and Unit\`a INFM\\
Mostra d'Oltremare, Pad.~19, I-80125 Napoli (Italy)}
\date{}
\maketitle

\begin{abstract}
These lectures contain a brief description of evolutionary models
inspired by the statistical mechanics of disordered systems.
After an introduction describing the Darwinian paradigm
of evolving populations, the deterministic quasispecies equation
is described, and the simplest fitness landscapes are discussed.
The effect of finite population size is then considered, from
the opposing points of view leading to stochastic escape and to
adaptive walks. A synthesis is attempted. Finally the effects of coevolution
are considered, and the promising models of large-scale
inspired by the Bak-Sneppen models are described.
\end{abstract}

\section{Introduction}
The subject of these lectures are some mathematical models of biological
evolution. For an elementary introduction to evolutionary
genetics one can look at ref.~\cite{MS}.
We shall see here that many concepts from the statistical physics
of disordered systems find their application in evolutionary biology,
what motivates the presence of these lectures in a workshop
dedicated to the dynamics of disordered and frustrated systems.

We shall first dwell on evolution at the level of a single
population ({\em microevolution\/}). In section 1 we introduce a rather
general description of the population dynamics: a population of living
individuals reproduces under the constraints imposed by limited resources. Every
individual passes to its offspring inheritable characters, on
which natural selection acts. Mutations affect the transmission of
inheritable information, creating new variability. We consider a static
environment, i.e., a fixed {\em fitness landscape\/} is assumed (cf.~\cite{Geilo}). 

Section 2 introduces the ``quasispecies" theory, which gives a
{\em deterministic\/} description of evolution, reminiscent of the equations
of chemical kinetics~\cite{Ei,ECS}. The quasispecies approach focuses on the
competition between random mutations and natural selection.
These two terms can be
put in formal correspondence with entropy and energy in a
thermodynamic system, and the evolutionary system
can be thus represented by a statistical mechanical model~\cite{Leu,BB}.
Transitions from an adaptive phase to a disordered neutral phase are
observed when the mutation rate crosses the ``error threshold"~\cite{ECS,Tara,PH}.

In section 3 the fluctuations of the reproductive process,
which take place in populations of finite size, are taken
into account and evolution is described as a stochastic process. In
such situations the adaptation level can go down ({\em Muller's ratchet\/}) or 
even disappear ({\em stochastic escape\/})~\cite{HW,WH}.

In section 4 we consider a higher level of modelling: the coevolution of
different species. In the models, species interact and the fitness of
the individuals of one species is shaped 
by this interaction. We first consider a model of the host-parasite
interaction, and we then turn to models that describe the evolution at
the level of the global ecosystem ({\em macroevolution\/}). 
The mathematical modelling of macroevolution has recently received
much attention by physicists, stimulated by new results about the statistical
properties of extinction events and by new theoretical perspectives.

\section{The Darwinian Paradigm}\label{Darw}
We start this series of lectures with a simplified but rather
general description of the evolution at the level of a single population
of reproducing individuals ({\em microevolution\/}). In this process, a
set of inheritable characters ({\em genome\/}) is passed from parent to
offspring. Random mutations and natural selection act on the
genomes. The term natural selection expresses the fact that different
characters have different reproductive potentialities {\em in a given environment}.

For simplicity, we deal with asexual reproduction, but a similar
framework can be used to describe the evolution of a sexual population.
The model we introduce is inspired by an algorithm for the
stochastic kinetics of coupled chemical reactions~\cite{Gill}
that can be easily implemented on a computer.
It is based on the following simplifications: 
\begin{description}
\item[Constant population:] The number $M$ of individuals does not change
  with time. The constraint
  of fixed population size models the struggle for life in
  an environment with limited resources. More general constraints do
  not change qualitatively the results. In several models, an infinite
  population is considered: this is a simplification that allows to
  neglect stochastic effects in the reproductive process. Models with
  finite populations often show features which do not appear in
  the infinite population limit. 
\item[Constant genome length:] The inheritable characters of
each individual are encoded in a
  string of $N$ symbols (for simplicity, and without loss of
  generality, we consider binary symbols), $s_i^\a=\pm 1$, where 
   $i=1,\dots, N$ labels the position in the sequence and 
   $\a=1,\dots, M$ labels the individual to which the genome
  belongs. $N$ is fixed, thus we do not consider the possibility that
  inheritable information is increased (or decreased) during
  evolution.Thus genome space is  represented by the
  $2^N$ vertices of the hypercube $\{-1,1\}^N$.
\item[Non-overlapping generations:] All the individuals in the
  population at generation $t$ are replaced by their offspring at
  generation $t+1$. This situation may happen in nature, for instance,
  in a wheat field, where each generation has a one-year
  span. With this assumption, time is a discrete
  variable measuring the number of generations.
\end{description}

With these assumptions, the state of the population at time $t$ can be
described by specifying the genomes of all the individuals,
$\{{\bf s}^\a(t)\}, \a=1,\dots, M$ (where ${\bf s}=(s_1,\dots,s_N)$) 
or, equivalently, by indicating, for each of the $2^N$ points
${\bf s}$ which make the  genome space, the number
$\nu_{\bf s}(t)$ of individuals with genome ${\bf s}$. 
Typically, most of these occupation
numbers vanish: biological populations are extremely sparse in genome
space. Typical orders of magnitude are 
\be 
N\approx 10^6\div 10^9 \ll M\approx 10^9 \div 10^{12}\ll 2^N.\nonumber
\ee 

The point of view we adopt is rather different from that of classical
genetics. There, attention is focused on the presence (or absence) of
few characteristic traits. These traits are governed by specific sites
({\em loci\/}) in the genome, where one of a few genetic variants 
({\em alleles\/}) may be found. The stress is laid upon the change
of the frequency of a given allele during the evolutionary process.
Since the alleles are few, it is warranted to assume that each of them 
is carried by a large number of individuals, and one can thus
apply the usual methods of probability theory. On the other hand, this
point of view leads almost without alternatives to a picture in which
different alleles struggle to increase their frequency at a given locus,
independently of what takes place at other loci. Only in a few
cases one is able to take into account the fact that the effect of
the presence of a given allele in a locus depends on what alleles are
present in some other loci (this effect is called {\em epistatic interaction\/}).
The resulting picture is often called ``bean-bag genetics'', as if
the genome were nothing else as a bag carrying the different alleles
within itself. The ``global'' point of view we adopt here aims at
providing at least a language in which the stage for the understanding
of the effects of epistatic interactions on the evolutionary behavior
can be set from the start. 

There are two other important simplifications that are used in the
dynamics of most microevolutionary models:
\begin{description}
\item[Constant environment:] The environment is not modified by the
  evolutionary process. In particular, the average rate of
  reproduction associated to a set of inheritable traits
  does not depend on the composition of the population. In
  other words, in such a situation there is no interaction between the
  individuals in the population, apart for the competition for resources.
\item[Constant mutation rate:] The mutation rate is independent of
the locus (i.e., of the unit of the genome one considers)
and is constant from generation to generation.
In particular, it is not considered to be itself subject to
  genetic control.
\end{description}

The evolutionary process can then be represented as
a three stage sto\-chas\-tic process:
\begin{enumerate}
\item {\bf Reproduction:} The individual $\a$ at generation $t$ is the
  offspring of an individual living at generation $t-1$. Reproduction
  is thus represented as a stochastic map
\be 
\a\longrightarrow \a'=G_t(\a), \ee
where $G_t(\a)$ is the parent of the individual $\a$, and is chosen at
random among the $M$ individuals living at generation $t-1$. 
\item {\bf Mutation:} The genomes inherited by all of the individuals in the
  population 
  undergo independent random changes. The assumption of a constant
  genome length simplifies the treatment of such process.
  A further simplification consists in considering only
  independent point mutation, i.e., every element of the genome
  is modified with a given probability independent of the other
  elements,  namely
\be
 s_i^\a(t)=-s_i^{G_t(\a)}(t-1)\qquad \hbox{with probability } \m, \ee
where the parameter $\m\in [0,1/2]$ is the microscopic {\em mutation rate}. In
real organisms, more complex phenomena take place, like
global rearrangements of the genome, copies of some part of the genome,
displacements of blocks of elements from one location to another
one\dots\  However, consideration of such correlated mutations makes the
model much more difficult to treat and does not add much insight, at
our rather abstract level of description.
\item {\bf Selection:} The expected number of offspring of each individual
depends on its genome, and is evaluated in this
  stage. It is proportional to a quantity called the {\em fitness\/} of
  the genome. 
\end{enumerate}

 This quantity is one of the most debated in
  population genetics since it was introduced by
 Ronald A. Fisher~\cite{Fi} and Sewall Wright~\cite{Wr}. 
 Its formal definition is the
  following:
  \begin{quote}
 The fitness of a {\em phenotypc trait\/} 
  is proportional to the average number of offspring
  produced by an individual possessing that trait, {\em in a given
  existing population}.
  \end{quote}
We remark that this notion of fitness is a concept defined at the level of
individuals in an homogeneous population, and it is difficult
at this point to
speak about the fitness of a species or of a group of species.

We are going to generalize the concept of fitness (which is related to a
single---or at most a few---phenotypic traits in a given population)
by associating it to the whole {\em genotype} ${\bf s}$. This is a rather
bold step, since the fitness such defined cannot be measured, due to
the fact mentioned above, that
most genotypes are not encountered in a given population.
We therefore adopt the following definiton of fitness:
\begin{quote}
The fitness of a {\em genotype\/} ${\bf s}$ is {\em proportional\/}
to the {\em average\/} number of offspring of an individual possessing
the genotype ${\bf s}$.
\end{quote} 
With this definition we have tacitly introduced an additional hypothesis,
namely that the reproductive success of an individual depends on its genotype
alone, up to a proportionality constant. In general, this is not true:
the reproductive value of a given trait can depend on its frequency in
the population (one can think at the effects of sexual selection, where
rare, but not too odd, traits often entail preference and hence reproductive
success). However this simplifying assumption is a good starting point.
The essential point of this definition is the consideration
of the {\em average\/} number of offspring instead of
the {\em actual\/} one. This reflects the intrinsically stochastic
nature of the reproduction process. As it is nicely put by 
John Maynard-Smith~\cite[p.~38]{MS}:
\begin{quote}
If the first human infant with a gene for levitation were struck by lightning in its
pram, this would not prove the new genotype to have low fitness,
but only that the particular child was unlucky.
\end{quote}

Since the fitness that we have defined is a nonnegative quantity,
we choose to represent it with the notation
\be 
\hbox{Fitness}({\bf s})=W({\bf s})=\E^{k F({\bf s})}
\propto \hbox{Average number of
  offspring}({\bf s}). \ee
The reason of the exponential representation of the fitness will be
clear in next section. The necessity of introducing an
(unspecified) proportionality constant stems from
the assumption of a constant population size, which
makes the reproductive success a {\em relative\/}
notion. It is easy to give sense to fitness {\em ratios\/}
(this genotype is twice more successful that that one,
because on average it has twice the number of offspring
than that one), but it is much harder to give it
to absolute values. It follows that the quantity $W({\bf s})$ is defined
up to a proportionality constant and, therefore, that
the function $F({\bf s})$ only up to an additive constant, much like an energy.
We have also introduced an inverse ``selective temperature'' $k$,
which shall turn useful later.

If we imagine to draw a line above each point ${\bf s}$ in genotype space, of height
proportional to $F({\bf s})$, we obtain what is called
a {\em fitness landscape}. We can imagine the evolutionary process
taking place in this landscape, each individual being represented
by a point on top of its genotype. The evolving population wanders
therefore on the landscape like a flock of sheep, and our first aim 
is to characterize its motion.

The earliest result concerning this problem is 
the so-called Fundamental Theorem
of Natural Selection, first stated by Fisher~\cite[Chap.~II]{Fi}. The theorem
says that, in the absence of mutations and in the limit of an infinite
population (so that the fluctuations of the reproductive process can
be neglected) the average fitness of the population cannot decrease
in time, and becomes stationary only when all of the individuals in the
population bear an optimal genome, corresponding to the maximum
value of the fitness.

We shall prove the theorem for the simpler case of asexual
reproduction (the original version is concerned with the sexual case,
which is much more complicated). The proof
runs as follows: we define
\be 
\l\la W\r\ra_t={1\over M}\sum_\a W\l({\bf s}^\a(t)\r)={1\over M}
\sum_{\bf s}W\l({\bf s}\r)\n_{\bf s}(t), \ee
as the average fitness of the population (angular brackets will
denote from here on population averages). The evolution equations, in
the above hypothesis (absence of mutations and deterministic asexual
reproduction) are given by
\be 
\n_{\bf s}(t+1)=\frac{1}{\l\la W\r\ra_t}\n_{\bf s}(t)W({\bf s}). \ee
The normalizing factor $1/\l\la W\r\ra_t$ is chosen so that
the population size remains constant:
\be
\sum_{\bf s}n_{\bf s}(t+1)=\frac{1}{\l\la W\r\ra_t}\sum_{\bf s}
\n_{\bf s}(t)W({\bf s})=M.\ee
Then
\be \la W\ra_{t+1} ={\sum_{\bf s} W({\bf s})^2\n_{\bf s}(t)\over\sum_{{\bf s}'}
  W({\bf s}')\n_{{\bf s}'}(t)} = {\l\la W^2\r\ra_t\over \l\la W\r\ra_t}\geq
  \l\la W\r\ra_t , 
\ee
where the equality applies only if all individuals bear an
  optimal genotype (i.e., a genotype corresponding to
  the maximum fitness).
  
This result was emphasized since the early days of population
genetics. A recent commentary by Karl Sigmund~\cite[p.~108]{Sigm} hints that
it should be taken {\em cum grano salis\/}:
\begin{quote}
So we see, in physics, disorder growing inexorably in systems isolated
from their surroundings; and in biology, fitness increasing steadily
in populations struggling for life. 
Ascent here and degradation there---almost too good to be true.
\end{quote}
In fact, it does not seem to be absolutely true. 
If such were the case, it will be hard to
understand the origin of the remarkable
variability of living beings, the variability that provides
the very material for the evolutionary process! On the
other hand, this view of
evolution as an everlasting improvement has recently met a deep crisis, both in
the microevolutionary context and in the broader context of the evolution
of ecosystems. In microevolutionary models, consideration of finite
populations and of random mutations shows that the increase in fitness
stated by the Fundamental Theorem holds just in particular situations,
and is apparently more the exception than the rule. In next section we shall
see how mutations change the picture of evolution in a deterministic
theory. In section 3 we will consider the effects of finite
population, that introduces stochasticity in the reproductive process.

\section{The quasispecies theory}
The quasi-species theory was introduced by Manfred Eigen in 1971 to describe
the evolution of a system of information carrying macromolecules through
a set of equations of chemical kinetics~\cite{Ei}. The equations are
deterministic (one assumes that population size is infinite), and
reproduction takes place asexually. Emphasis is laid on the competition
between natural selection and random mutations.

We introduce normalized population variables,
\be 
x_{\bf s}(t)={\n_{\bf s}(t)\over M}. \ee
Since the population is infinite, the actual number of
offspring of an individual bearing a genotype ${\bf s}$
is proportional to its {\em expected\/} value, and therefore
to its fitness $W({\bf s})$.
The evolution equations are therefore
\be x_{\bf s}(t+1)={\sum_{{\bf s}'}x_{{\bf s}'}(t)W({\bf s}') 
Q_\m\l({\bf s}'\ri {\bf s}\r)
\over \sum_{{\bf s}'}W_{{\bf s}'}x_{{\bf s}'}(t)}. \ee
We have introduced the mutation matrix $Q_\m\l({\bf s}'\ri {\bf s}\r)$
(dependent on the mutation rate $\m$) whose elements
are the conditional probabilites that, in the attempt of
reproducing an individual with genotype ${\bf s}'$
one obtains a genotype ${\bf s}$. As we discussed in the previous
section, we consider a very simplified mutation pattern: the genome
length is kept constant, and only point mutations are allowed at every
location, independent of one another. In this case, the mutation
probability  $Q_\m\l({\bf s}'\ri {\bf s}\r)$ depends only on the Hamming distance
$d_{\rm H}$ between ${\bf s}$ and ${\bf s}'$, i.e., on
the number of units that are
different in the two configurations:
\be d_{\rm H}({\bf s},{\bf s}')=\sum_{i=1}^N {(s_i-s'_i)^2\over 4}. \ee
One has
\be 
Q_\m\l({\bf s}'\ri {\bf s}\r)=\m^{d_{\rm H}}(1-\m)^{N-d_{\rm H}}
\propto \exp\l(-\b\sum_i
s_is_i'\r), \ee
where $\b$ is defined by
\be 
\b={1\over 2}\log\l(1-\m\over \m\r). \ee
The notation anticipates the analogy between the mutation
coefficient $\b$ and the inverse temperature in a thermodynamical
system. Using the exponential representation of the reproduction
weight $W$, we can write the evolution in a form
that is suggestive of a statistical mechanics analogy:
\be 
x_{\bf s}(t+1)={1\over \l\la W\r\ra_t}\sum_{{\bf s}'} x_{{\bf s}'}(t)
\exp\l(\b\sum_is_i s'_i+kF({\bf s}')\r). \label{map}\ee

It is worth remarking that these equations are
non-linear in the dynamical variables $x_{\bf s}(t)$ only because of the
normalization condition. It is thus convenient to introduce the 
unnormalized variables
$y_{\bf s}(t)$ that satisfy linear equations of motion:
\be 
y_{\bf s}(t+1)=\sum_{{\bf s}'} y_{{\bf s}'}(t)
\exp\l(\b\sum_is_is'_i+kF({\bf s}')\r). \label{map2}\ee
The relation between the $y_{\bf s}$'s and the $x_{\bf s}$'s,
stems from the normalization condition imposed on the $x_{\bf s}$'s:
\be
x_{\bf s}(t)=\frac{y_{\bf s}(t)}{\sum_{{\bf s}'}y_{{\bf s}'}(t)}.\ee
Equation (\ref{map2}) reminds one of the solution of
a statistical mechanics model via the transfer matrix formalism.
In fact, it is possible to map
the time evolution into a statistical mechanics problem
in a two-dimensional space, where the two coordinates represent time
and genome coordinate~\cite{Leu,BB}. The effective Hamiltonian is given by
\be 
\b H= \b\sum_{i,t}s_i(t) s'_i(t+1)+k\sum_t F({\bf s}'(t)). \ee
Formally, the situation is similar to a model
of Quantum Spin Glass. We are interested in the asymptotic state of
the system, which correspond to the last time layer. Thus the
evolutionary problem corresponds to a surface problem. 

Several fitness landscapes have been studied in the literature. As
an example, we consider here two extreme cases of a landscapes with a
single peak: a very smooth
fitness landscape, sometimes called the Fujiyama
landscape, and
a very rugged landscape where there is a single isolated peak
surmounting a sea of equivalent low fit genotypes.

In the first one,
the fitness increases regularly toward the peak in all
directions, and walking in this landscape is
like climbing a smooth volcano, in the sense that at
any point it is possible to point directly to the top
by climing in the direction of maximal slope. It is defined by
$F({\bf s})=\sum_ih_is_i$ (without any essential
loss of generality, we put $h_i=1$).

In the second one, one has  $F({\bf s})\propto \d_{{\bf s}{\bf s}_0}$.
In other words, all genotypes have the same fitness value,
except one (the ``master'' or ``preferred'' genotype) that has a higher value.
This landscape is often called the sharp peak landscape, and
has apparently been introduced by John Maynard-Smith in 1983,
although I have been unable to locate the reference. 
Here the situation is the opposite: it is not possible
to know where the fitness top is, unless one
is exactly on it! We shall
show that in the second case the quasispecies model undergoes a
transition between
an adaptive regime, where evolution is ruled by selection,
and a neutral regime, where the evolution is essentially driven by
random mutations, and that this transition can be described analogously to a phase
transition in equilibrium statistical mechanics~\cite{Leu,Tara}.

\subsection{The Fujiyama landscape}
This landscape, $F({\bf s})=\sum_is_i$, is characterized by the absence of
interactions between genome elements. In this case the statistical
mechanics terminology and the genetic terminology agree:
genetists call this one the landscape ``without epistatic
interaction" (the term epistatic, that sounds somehow obscure to
non-genetists, refers to the interactions between
different genes).

We address here the question of the limit distribution of the population
in the genome space, described by
\be 
x^*_{\bf s}=\lim_{t\ri\infty} x_{\bf s}(t), \ee
that is independent of the initial distribution (in the conditions
where the infinite size limit of the corresponding statistical
mechanical model exists). As it was suggested above, we shall use the
variables $y_{\bf s}$, whose evolution is governed by a linear equation.

It is easy to see that, due to the absence of interactions, if in the
 initial state there are no ``correlations" in genome space (i.e.,
 $y_{\bf s}(0)=\prod_i y_{s_i}(0)$), the genome elements will remain
 uncorrelated forever. A more detailed analysis shows that
 even if such initial correlations are present, they are
 broken up after a number of generations which depends on
 $\b$. Thus the asymptotic state does not exhibit correlations:
 \be
 x^*_{\bf s}=\prod_i x^*_{s_i}.\ee
 It is therefore enough to study the dynamics of a single genome
 unit, say $s_i$:
\be 
y_{s_i}(t+1)=\sum_{s_i(t)}y_{s_i}(t)\exp \l(\b
s_i(t)s_i(t+1)+ks_i(t)\r). \ee

The statistical mechanics analog of this evolution is a one-dimensional
Ising model with ferromagnetic
interactions in the time direction, an inverse temperature equal to
$\b$ and a magnetic field $k/\b$.  It is well known that such a
model does not have phase transitions. Thus we reach the conclusion
that evolution in the Fujiyama landscape takes place in a single
phase, where there always is some degree of adaptation.
One can evaluate it by introducing the ``order parameter''
\be
m=\frac{1}{N}\sum_i\left<s_i\right>,\ee
which is proportional (in our situation) to the average fitness.
One obtains~\cite{Tara}
\be
m=\sinh k\,\tanh \b\,\frac{\E^{\b}\sqrt{\E^{2\b}\sinh^2 k+\E^{-2\b}}
+\E^{2\b}\cosh k}{\E^{\b}\cosh k\sqrt{\E^{2\b}\sinh^2 k+\E^{-2\b}}
+\E^{2\b}\sinh^2 k +\E^{-2\b}}.\label{Fuji}\ee
The origin of the factor $\tanh \b$ is interesting. When one considers
the value of $m$ at generation $t$, one takes into account
the effects of selection (and therefore of $k$) only up
to generation $t-1$, and only the effects of mutation from generation
$t-1$ to $t$. This corresponds to a one-dimensional Ising model
in which the field (of intensity $k/\b$ is applied to all sites
by the last. Solving this problem by the transfer matrix method
yields eq.~(\ref{Fuji}).
We see therefore, whenever $k>0$, there is some degree of
adaptation for any nonzero value of $\b$, i.e., for
any mutation rate $\mu$ smaller than 1/2. As we shall soon see, this
conclusion is quite peculiar of this fitness landscape: epistatic
interactions introduce in the model a phase transition to a
non-adapting regime as soon as the error threshold is crossed.

\subsection{The sharp peak landscape}
This is a limiting case of very strong epistatic interactions: in this
case, any single element of the genotype does not give any 
information on the value of the fitness. 
This landscape is defined by the equation $F({\bf s})=\eps
\d_{{\bf s}{\bf s}_0}$. We shall treat it in the infinite genome limit,
$N\to\infty$, introduced by Kimura (see \cite[p.~236ff]{Ki}), and analogous
to the thermodynamical limit in statistical mechanics.
In order to have a nontrivial limit, we set $\eps=kN$. 
The dynamic equations then read 
\be
y_{\bf s}(t+1)=\sum_{{\bf s}'}y_{{\bf s}D'}(t)
\exp\l(\b\sum_is_is'_i+kN\d_{{\bf s}{\bf s}_0}\r). \ee

It is actually more transparent to consider
(following \cite{PH}) finite fitness for the master sequence
${\bf s}_0$ and a mutation rate with vanishes for $N\to \infty$
in such a way that the expected number of mutations
for each reproduction event is finite. We then define $x_k$ as 
the fraction of the population whose genotype has a Hamming distance
(``is $k$ mutations away'') sfrom the preferred genotype:
\be
x_k(t)=\frac{1}{M}\sum_{\bf s}\delta_{d_{\rm H}({\bf s},{\bf s}_0),k}\nu_{\bf s}(t).
\ee
The fitness $W({\bf s})$ is then given by
\be
W({\bf s})=\cases{1,&if ${\bf s}={\bf s}_0$,\cr
1-\s,&otherwise.}\ee
We take the $N\to\infty$
limit keeping $u=\m N$ finite, so that only a finite number
of mutations appear: if $u\ll 1$, as we shall assume, we can neglect
the possibility that multiple mutations appear.
We can moreover neglect, in the infinite genome limit, back
mutations that reduce the value of $k$, since they
have a probability proportional to $k/N\ll 1$. 
Thus we have two parameters, $u$
that measures the mutation rate and $\s$ that measures the strength of
the selection. Our approximations lead to the following evolution equations
\cite{PH}:
\bea x_0(t+1)&\propto& x_0(t)\l(1-u\r); \\
x_1(t+1)&\propto& u x_0(t)+\l(1-u\r)\l(1-\s\r)x_1(t); \\
x_k(t+1)&\propto& \l(u x_{k-1}(t)+(1-u)x_k(t)\r)(1-\s),
\qquad k>1. \eea
To normalize the $x_k$'s, we divide the r.h.s.'s by
the average fitness of the population, $\la W\ra=1-\s(1-x_0)$. We look
for the stationary distribution $\{x_k^*\}$. The equation for $x_0^*$
 does not involve the $x_k$ with $k>1$,
 in our approximations, and reads
\be 
x_0^*={x_0^*(1-u)\over 1-\s(1-x_0^*)}=\cases{1-u/\s, &if $u<\s$,\cr
0, &if $u\geq\s$.}\ee

We can thus distinguish two regimes: if $u<\s$,
one has $x^*_0>0$ and in fact (as we shall shortly see)
the whole population lies a finite distance away
from the preferred genotype. In this {\em adaptive regime\/}
the population forms what Eigen calls a {\em quasispecies},
i.e., a population of genetically close, but not identical
individuals. When $u>\s$, we have $x_k^*=0$, $\forall k$.
In this case, a closer look at the finite genome situation
shows that the population is distributed in an essentially
uniform way over the whole genotype space. The infinite
genome limit becomes therefore inconsistent, since
the whole population lies an infinite number of mutations
away from the preferred genotype. In this {\em wandering regime\/}
the effects of finite population size are prominent, and they
can be studied by using the concepts forged by Kimura
and the tenants of the Neutral Theory of molecular
evolution~\cite{Ki}. The transition from the adaptive (quasispecies)
regime to the wandering one is called the {\em error threshold}, 
and it is a quite generic feature of quasispecies theory.

To describe the transition in the statistical mechanics language, it
is convenient to define the overlap between two sequences:
\be q({\bf s},{\bf s}')={1\over N}\sum_{i=1}^N s_is'_i=1-{2d_H({\bf s},{\bf s}')\over
  N}. \label{overlap}\ee
The average overlap between the genomes in the population and the
master sequence can be used as an order parameter. It is the analogous
of a magnetization, $m=1/N\sum_i\la s_i\ra$. It is of order $1/N$ in
the neutral phase, while it is of order $1-O(1/N)$ in the adaptive
phase, so that it makes a finite jump at the transition~\cite{FP}.
A detailed solution of the quasispecies model in the sharp peak 
landscape has been recently obtained by S. Galluccio \cite{Gall}.

\subsection{Rugged fitness landscapes}
We have seen that the sharp peak landscape exhibits a ``phase transition'',
the error threshold, which does not take place in the Fujiyama landscape. It
is interesting to interpolate between these two extreme situations.
A relevant quantity under this respect is the {\em
ruggedness\/} of the fitness landscape. This quantity plays a very
important role in determining the qualitative features of the
evolution, as it was pointed out by P. W. Anderson~\cite{An} and
more systematically by Kauffman
\cite{KL,Ka}, who introduced a one-parameter family of fitness landscapes of
increasing ruggedness. Our definition is slightly different from the
original one by Kauffman, and coincides with the $K$-spin Hamiltonian,
familiar in the context of disordered systems:
\be 
F_K({\bf s})=\sum_{\{i_1\ldots i_K\}} J_{i_1\ldots i_K}s_{i_1}\cdots
s_{i_K}. \ee
Here the $J_{i_1\ldots i_K}$, for each different set
of indices $\{i_1,\ldots,i_K\}$, are independent, identically
distributed, random variables, so that
for every $K$ we are dealing with a random ensemble of fitness
landscapes. The variance of the $J$'s is chosen in
a way that guarantees a meaningful infinite genome 
limit:
\be
\Delta J^2=\left[J_{i_1\dots i_K}J_{j_1\dots j_K}\right]_{\rm av}
= \frac{N^{-K+1}}{K!}\prod_{\a=1}^K \d_{i_{\alpha}j_{\alpha}}.
\ee
We denote here by $\left[\ldots\right]_{\rm av}$ the average
taken over all possible realizations of the random variables $J$.
The larger $K$, the faster the fitness
correlations decay in sequence space, so that the fitness
landscape is less and less correlated, i.e., as it is usually
said, more and more rugged:
\be \left[F_K({\bf s})F_K({\bf s}')\right]_{\rm av}= K!N^K{\Delta J}^2
q({\bf s},{\bf s}')^K=N  q({\bf s},{\bf s}')^K.\ee

In the $K\to\infty$ limit, one has $\left[F_K({\bf s})F_K({\bf s}')\right]_{\rm av}
=N \delta_{q({\bf s},{\bf s}'),1}$. Thus
 this fitness landscape coincides the Random Energy
Model hamiltonian introduced by Derrida
in the theory of spin glasses~\cite{D},
where the values of the fitness at different
positions in sequence space are independent random variables. In the
genetic literature, this limit is often referred to as the {\em rugged
fitness landscape}. We give a brief description of the quasispecies
model in a rugged fitness landscape (cfr.~\cite{FPS}). 
The $F({\bf s})$ are independent Gaussian variables. In order to obtain
a non-trivial infinite genome limit, their variance has to be
proportional to $N$: we choose $\left[F({\bf s})^2\right]_{\rm av}=N/2$. 
We imagine that, at time $t$, the
population is located  on the highest peak of the fitness landscape,
corresponding to $F({\bf s})=E^*$.
The average number ${\cal N}(E,q)$
of sequences with
$F({\bf s})=E$ and whose overlap with a given one is equal to $q$ is given by
\be 
{\cal N} (E,q)\simeq \exp\l(N{\cal S}(q)-E^2/N\r), \ee
where ${\cal S}(q)$ is obtained from the Stirling formula for the binomial
coefficient:
\be 
{\cal S}(q)=\ln 2-{1\over 2}\l[(1+q)\ln (1+q)+(1-q)\ln(1-q)\r]. \ee

We can then distinguish two cases.
If ${\cal N}(E,q)$ is large, we can identify it with the {\em typical\/}
number of sequences. In this hypothesis,
the partition function of the corresponding statistical mechanical
model from time step $t$ to time step $t+1$  reads
\be 
{\cal Z}\approx \int_{({\bf s}(q)-E'^2)>0} \frac{\D q \D E'}{N }\exp\l(N(kE'+\b
q+{\bf s}(q)-E'^2\r), \ee
and the integral can be evaluated with the saddle-point method. 
Thus, the main contribution stems from
the maximum at $E'=k/2$, $q=1-2\m$.
But there is also a situation where the main
contribution comes from the highest peak: $q=1$, $E=E^*$.
The typical value of the optimal fitness $E^*$ can be obtained from the
condition $\exp\l(N\ln 2-(E^*)^2/N\r)=O(1)$. From this follows
$E^*=\sqrt{\ln 2}$. Comparing the two values of the ``free energy",
$1/N\ln {\cal Z}$, we find that the transition takes place at
\be 
k_{\rm c}=2\l(\sqrt{\ln 2}-\sqrt{\ln 2+\ln(1-\m)}\r). \ee
We have thus obtained two phases: the frozen phase,
where, at each generation, only individuals possessing
an optimum genotype (with $F=E^*$) can reproduce;
and the free (or wandering) phase, in which the effects
of mutations rapidly overcomes that of selection.
Locally the sharp peak landscape picture holds. 
One can define an order parameter by considering
the average overlap of the population
with itself over a very large time
span (this is known is spin glass theory as the
Edwards-Anderson order parameter):
\be
q_{\rm EA}=\lim_{t\to\infty}\lim_{t'\to\infty}\frac{1}{N}
\sum_i
\left[\left<s_i(t)\right>\left<s_i(t')\right>\right]_{\rm av}.
\ee
This order parameter drops from a finite value ($\tanh\beta$)
to zero as one goes from the frozen to the wandering
phase, crossing the error threshold.
One can also the $K=2$
landscape. This case is formally related to the Quantum Spin Glass model
discussed by A.P. Young in this College, and also predicts an error
threshold.

\section{Finite populations}
The quasi-species model is inconsistent in the neutral regime. In
fact, the population is in this case spread in genome space, and the
infinite population limit is not reasonable anymore. In this
situation, the fluctuations of the reproductive process in a finite
population have to be taken into account. We briefly recall the
notation and the stochastic dynamical rules introduced in section
\ref{Darw}:$\a\in\{1,\ldots,M\}$ labels
the individuals in the population, and  ${\bf s}^\a(t)=(s_1^\a(t), \ldots
s_N^\a(t))$, $ s_i=\pm 1$ represents the genome of the
individual $\a$ through a sequence of $N$ binary symbols. 
At each generation $t$, the following generation is
obtained in two steps:
\begin{enumerate}
\item Reproduction: For every $\a\in\{1,\ldots,M\}$, we extract the parent
  $\a'=G_{t+1}(\a)$ at random, with probability
\be 
\Pr \l\{G_{t+1}(\a)=\b\r\}={W\l({\bf s}^\b(t)\r)\over \sum_{\b'=1}^M
  W\l({\bf s}^{\b'}(t)\r)} . \ee
\item Mutation: Independently from one another, the genome elements
can change respect to those of the parent:
\be 
\Pr\l\{s_i^\a(t)=-s_i^{\a'}\r\}= {1\over
  2}\l(1-\E^{-2\m}\r), \ee
where $\m$ is the mutation rate, defined in a slightly different way
with respect to section \ref{Darw}. This definition will turn out to be
  more convenient in the following. At the first order in $\m$, the mutation
  probability is simply $\m$ and the two definitions coincide.
\end{enumerate}

We start considering a flat fitness landscape, $W({\bf s})=\hbox{const.}$: this
means that all genomes are equivalent and natural selection does not
act. This case would be trivial in the deterministic model, but it is
interesting for a finite population. In this case, the constraint of
limited resources, that we implemented as the constraint of a finite
and constant number of individuals in the population, does produce an order in
sequence space even in the absence of natural selection~\cite{DP}.

This order can be studied through the distribution of the overlap
(\ref{overlap}) in the population. Formally, this is defined as
\be P(q)=\l\la \d\l(q({\bf s}^\a,{\bf s}^\b)-q\r)\r\ra. \ee
The labels $\a$ and $\b$ identify the individuals, and the
angular brackets mean a population average. However,
this quantity fluctuates from generation to generation, and
it is necessary to consider its average over all possible realizations
of the reproduction process. This situation recalls the need for
disorder averages on top of themal averages in the theory
of disordered systems. We shall denote this
average by a bar: $\overline{\mathstrut \ldots}$

In the infinite genome limit,
the overlap $q({\bf s}^\a,{\bf s}^\b)$ is directly related 
to the number of generations
passed since the last common ancestor of individuals $\a$ and $\b$
was living. This quantity, $\t_{\a\b}$, is a measure of
distance between the individuals in the population.
Not only $\t_{\a\b}$ is a distance, under the conditions of asexual
reproduction, but it can be shown to be {\em ultrametric},
since it satisfies the inequality
$\t_{\a\b}\leq \max\l(\t_{\a\g},\t_{\b\g} \r)$. 
This property is crucial for the taxonomic ordering of the
population into clusters of individuals with closer common origin.
The relation between $q({\bf s}^\a,{\bf s}^\b)$ and $\t_{\a\b}$ is very simple:
$q$ is the correlation between the initial and the final state after
a random walk due to mutations, lasting $2\t_{\a\b}$
generations. Indicating with the symbol $\l[\ldots\r]_{\rm mut}$ the
average taken over the mutation process, we have
\be \l[q({\bf s}^\a,{\bf s}^\b)\r]_{\rm mut}=\exp(-4\m\t_{\a\b}), \label{q}\ee
 In the infinite genome limit, the fluctuations of $q_{\a\b}$
vanish and the above relation can be taken to be a deterministic
relation between $q_{\a\b}$ and $\t_{\a\b}$. Thus the distribution of
$q$ gives interesting informations about the taxonomic structure
of the population: indeed, in modern taxonomy, the genetic similarity
between contemporary species is more and more used to reconstruct
taxonomic trees.

It is interesting to look at the snapshots of $P(q)$ at different generations~\cite{HD}. 
This is a very broad distribution, with many peaks that move in
time. The height of the peak is related to the size of the cluster in
the population whose last common ancestor lived $-1/(4\m\log q)$ generations
ago. The  large peaks move towards $q=0$ following an exponential law: 
$q\propto \E^{-4\m
  t}$. They  represent the common ancestors of large clusters in the
populations. At the same time, small peaks are continuously created at
large $q$, and eventually increase in size while they shift towards
$q=0$. Thus the distribution looks completely different from one
snapshot to another one, even in the infinite genome limit. 
In the language of disordered systems, one could say that $P(q)$ 
is not self-averaging.
It is noteworthy that the $P(q)$ coming from a process of asexual
reproduction does show some important features of the order
parameter distribution function $P(q)$ defined
in  spin glass models, namely ultrametricity and lack of
self-averaging.

When we average $P(q)$ over
different realizations of the reproductive process (or, equivalently,
over time), we obtain a time-independent quantity $\overline{P(q)}$. 
It is not difficult to
compute this quantity in our model. We just have to compute the
distribution of $\t_{\a\b}$, the number of generations since when
the common ancestor of individuals $\a$ and $\b$ was living. To this
purpose, we imagine to follow the stochastic map $G_t(\a)$ backwards in the
past. What is the probability that, starting from two different individuals
$\a$ and $\b$ at generation $t$, $\t$ repeated applications of the map
$G_t$ still result in two different individuals at generation $t-\t$?
This probability, that we call $\n_\t$, is simply given by
\be \n_\t=\l(1-{1\over M}\r)^\t\simeq {\rm e}^{-\t/M}. \ee
This result shows that the last common ancestor of any two
individuals was living at most $O(M)$ generations ago.
The probability that the last common ancestor of two
individuals was living $\t$ generation ago is equal to $(1/M)\n_\t=(1/M)
\E^{-\t/M}$. From this result, it is easy to derive the distribution
for $q_{\a\b}$, considering that, in the infinite size limit, the
relation between the two variables is simply
$q_{\a\b}=\exp(-4\m\t_{\a\b})$. The probability density of this
variable is thus
\be 
\overline{P(q)}=\lam q^{\lam-1} \theta(q), \ee
where $\lam=1/4\m M$ gives a measure of the concentration of the
population in sequence space. In the limit $\lam\to \infty$ the
distribution is a $\d$ distribution in $q=1$. In the opposite limit
$\lam\to 0$ the distribution is a delta in $q=0$, which means that
the population has not anymore a structure and it is uniformly spread
in sequence space. It is worth noting that an ordering of the
population in sequence space ($\ov{\la q\ra}\neq 0$) is still present,
even in the absence of natural selection, 
if $\m=O(1/M)$.

Apart for the limiting cases $\lam=0$ and $\lam\to\infty$, the
overlap distribution has a finite width. Thus the population average
of the overlap, $Q=\la q_{\a\b}\ra$, is a non-self-averaging random
variable, whose process fluctuations do not vanish even in the
infinite population limit (if $\lam$ remains finite in this limit).

As the last argument concerning the flat landscape, we study
the dynamics of the population in sequence space. We consider the
average genome of the population as time $t$: $\la {\bf s}\ra_t=\l\{\la
s_i\ra_t\r\}$. We want to compute its autocorrelation function. It can
be easily seen that it decays exponentially to zero:
\be 
{1\over N}\sum_{i=1}^N\la s_i\ra_t\la
s_i\ra_{t+\t}=\overline{\left<q\right>}{\rm e}^{-4\m\t} \ee
(the proof is left to the reader as an exercise). The interesting
aspect of this formula is that the mutation rate for the population as
a whole is exactly equal to the mutation rate for a single individual,
$\m$ (in particular, it does not depend on the size of the
population). This is not intuitive for a
physicist, who is accustomed to think that a system made of many
objects moves slower than an isolated object.
The equality between
the ``macroscopic" mutation rate and the ``microscopic" one is an
important result of the neutral theory of molecular evolution,
developed mainly by Kimura. 
This consequence of the neutral hypothesis is very important
for the reconstruction of taxonomic trees from the observed genetic
similarity between extant species.
The neutral theory states that the extant genetic
data are in agreement with the hypothesis that most of the genetic
changes at the molecular level are due to selectively neutral
mutations. This is not necessarily in contradiction with
the evidence of adaptation. It only requires that the number of
selectively relevant traits is much smaller that the total number of
traits. In most globular proteins, for example, one can identify
a few aminoacids which are essential for function (and are therfore
strongly conserved by evolution) while most others can be substituted
(to a certain extent) without hindering the working of the protein.
The rate of aminoacid substitution in the latter ones  compares well
with that of {\em pseudogenes}, i.e., of those non-coding genome sequences wshich
are strongly correlated with those of existing enzymes, and are
believed to be non-functional copies of a working gene.

\subsection{The sharp peak landscape}
We now briefly discuss (following~\cite{HW})  the
evolution of a finite population in the two simple landscapes
considered in the framework of the quasi-species theory: the sharp
peak landscape and the Fujiyama landscape.
The first one is defined by the reproductive
strengths $W({\bf s}_0)=1$, $W({\bf s})=1-\s, 
{\bf s}\neq {\bf s}_0$. Let us define $M_0(t)$ as
the number of individuals whose genome is ${\bf s}_0$. This quantity follows
a Markovian stochastic process, with transition probability
\be 
\Pr\l\{M_0(t+1)=m'\mid M_0(t)=m\r\}={M\choose
  m'}\l(p_m\r)^{m'}\l(1-p_m\r)^{M-m'}, \label{proc}\ee
where, neglecting back mutations from a mutant genome towards the
  master sequence (we are considering the infinite genome limit), the
  parameter $p_m$ is given by
\be 
p_m={(1-\m)m\over m+(1-\s)(M-m)}. \ee

One can easily convince oneself that the asymptotic distribution is given by
$P(m)=\d_{m0}$: ultimately, a fluctuation will eliminate all copies of the
master sequences from the population, and no back mutation will be
able to restore them, no matter how large is the selective advantage
$\s$. Solving numerically equation (\ref{proc}) it is possible to
observe the transient behavior. Starting from a low concentration of master
sequences, at first the distribution
$P_t(m)$ moves towards larger $m$ values (the average number of
master sequences increases), but at the same time its width shrinks,
while the isolated peak at $m=0$ increases. The time scale at which the
width of the distribution for $m\neq 0$ vanishes depends on $M$, and
diverges, in the infinite population
limit, below the error threshold. Another signature of the
existence of an adaptive phase in the infinite size limit is the fact
that realizations of the stochastic process $M_0(t)$, starting from
the same initial condition, are self-averaging above the error
threshold (the fluctuations vanish in the $M\to\infty$ limit), while
they are not self-averaging below the threshold.

The phenomenon of the ultimate loss of the master sequence in a finite
population has been named {\em stochastic escape}.

\subsection{The Fujiyama landscape}
Here we consider a finite population in the Fujiyama landscape (Higgs and
Woodcock, 1995). We will see that new features appear: in contrast to
the deterministic case, where no error threshold transition takes
place in this particular landscape, a finite asexual population is not
able to occupy the optimal sequence even with very strong selective
advantage, if the mutation rate is finite. This phenomenon is known in
the genetic literature as ``Muller's ratchet".

Using a different parameterization, we define the Fujiyama landscape (no
epistatic interactions) through the equation
\be 
W_n=(1-\s)^n , \ee
where $W_n=\exp(k F_n)$ is the reproductive strength of the genomes
that are $n$ mutations away from the master sequence. The mutation rate per
individual and per generation is $u$. We will see that a
population that seats at the peak at $n=0$ will ultimately lose all of
the optimal sequences, no matter how large the selective advantage $\s$
is. This happens because the infinite genome limit has been
taken first, so that no back mutations towards the optimal genotype
take place. In this case, if the optimal genotype is lost for some
fluctuation in the reproductive process, there is no chance to get it
again. At this point the best genome in the population is one mutation
away from the master sequence, and the same reasoning can be applied
to it. The population is driven away from the peak by this stochastic
mechanism, where the fact that better mutations are vanishingly rare
acts as a ratchet (Muller's ratchet).
Thus, starting from $n=0$ as the initial state of the
population, we find that 
\be \la n\ra_t\approx Rt.\ee
  
In the limit $\s\to 0$ we find the flat landscape result,
which in this language reads $\overline{\la n\ra_t}=ut$ (process
average is needed, since $\la n\ra_t$ is a non-self-averaging
quantity in the neutral case). In other words, the mutation rate of
the population $R$ is equal to $u$ and does not depend on the
population size. On the other hand, as soon as $\s>0$ the mutation rate 
$R$ vanishes as $M\to\infty$, as
we know from the quasispecies theory. 

As we said, these results hold if the limit $N\to\infty$ is taken
first. If the genome is finite, the probability
of advantageous mutations cannot be neglected, and the
population ends up hovering at some average distance $\la
n\ra^*$ from the master sequence, which depends in a complicated
way on $u$, $s$, $M$ and $N$.

We started this overview of microevolutionary models with the
``optimistic" point of view of Fisher, according to which the fitness of
a population is a quantity that can not decrease in time. This is,
according to him, the main feature of the evolutionary process.

Then we considered non-vanishing mutation rate in very large
populations, and we discovered the error threshold transition: above a
given mutation rate, the increase of fitness is {\em not\/} the driving
force of evolution. In this case, the deterministic description is not
valid anymore, and we are forced to consider the fluctuations of the
reproductive process in finite populations. In this way we learned that,
even below the error threshold, fitness may indeed {\em decrease\/} in a
finite population. Which situation is most common in nature? According
to Kimura's neutral theory, most of the genetic changes at the
molecular level have been produced by selectively neutral
mutations. This neutral hypothesis is still rather vehemently discussed.

We are not going into this dispute, but we discuss briefly an experiment
and a model concerning a viral population that resurrect the
``optimistic" point of view of the increase of fitness. In this experiment,
 some viruses infect a cellular culture. After a
given time, a probe of the viral population is transmitted to another
culture. Fitness is measured as the spreading speed of the
viruses, compared with that of a control culture,
and it is observed to increase monotonically in time (in the
first 100 transmissions), with a tendency to exponential increase at
long times \cite{No}. A model that reproduces
very well the experimental data is based on a one-dimensional fitness
landscape \cite{Ts}. Reproduction is deterministic
and mutations are modeled as diffusion in this one dimensional fitness
landscape, whose coordinate is the reproductive rate $w$. At the
mean-field level, the model is described by the equation
\be 
{\pa p(w,t)\over\pa t}=\th\l(p-p_c\r)(w-\la w\ra_t)p(w,t)+D{\pa
  p(w,t)\over \pa w^2}, \ee
where $\la w\ra_t$ is the average fitness of the population.
However, this equation predicts that the fitness of the
populations goes to infinity in a finite time! The paradox is
solved by taking into account the effects of finite population size.
As a results, the authors of ref.~\cite{Ts} find that $\la w\ra_t$
increases linearly with time, with a rate which depends in a complicated
way on population size and mutation rate.

A more general analysis of both this experiment and the phenomenon of
the ratchet is possible: one can ask what is the rate of
accumulation of mutations in a finite population evolving in a smooth fitness
landscape $W_n=(1-\s)^n$ with mutation rate $u$, if the fraction
of favorable mutations is $p$ \cite{WH}. It is found
that, while for $p$ smaller than a threshold $p^*\approx 0.11$
disadvantageous mutations are accumulating at a rate $R$ increasing
with $u$, for $p>p^*$ and small $u$ there is a
regime where favorable mutations accumulate at a rate also increasing
with $u$. Thus both the experiment and the model described above may
be interpreted as representing this situation. It is likely, however,
that the fraction of favorable mutations is very small in most
realistic biological situations (for instance, if viruses had been
 able to reproduce at an ever increasing rate, we would have gone
extinct long ago!)

\subsection{Adaptive walks}
A more coarse-grained description of population dynamics has been
proposed in the literature \cite{KL,Ka}. The population is
represented by a single point in genome space (the genomes of all
individuals are considered equal). One assumes that
the population is finite, the selective pressure is very strong and
the mutation rate is small. Under these hypotheses, one can describe the
dynamics in the following way: at each time step, only one genome element of
some individual in the population mutates. If, because of this mutation,
one obtains a genotype with higher fitness, the new genotype
spreads rapidly through the entire population,
that moves therefore to the new position in genome space. If the
fitness of the new genotype is lower, the mutation is 
rejected and the population remains in the
old position. This process leads therefore to a local fitness optimum.

Physicists would call this process a Monte-Carlo dynamics at zero
temperature. As it is well-known, this algorithm does not lead
to a global optimum, but to a ``typical" local optimum. It is thus
important to investigate the statistical properties of the local
optima. One finds that these properties depend strongly on the
ruggedness of the fitness landscape,as parameterized, e.g., by the
parameter $K$ in the $NK$ landscapes introduced by
Kauffman. In the limit of extreme ruggedness 
there are no correlations between the values of the fitness at
any two different locations in genome space and the landscape
coincides with the Random Energy Model. In this case, many
quantities of interest can be computed analytically.

Let us consider this case. Let us denote by $N$ the number of genome
elements and by $F$ the fitness,
uniformly distributed between 0 and 1 (if we were considering a
fitness $F'$ distributed with a density $\rh(F')$, we would recover the
previous case through the transformation $F=\int_0^{F'}\rh(x) \D x$). 
The probability that a point with fitness $F$ is a local optimum is
simply given by $F^N$, since we have to impose that the $N$ nearest
neighbors of the point have fitness less than $F$. The probability
that a point is a local optimum is given by
\be
\Pr\l\{\hbox{local optimum}\r\}=\int_0^1F^N \D F={1\over N+1}. \ee
There are therfore a great deal of local optima. At every successful step
the distance from the top is divided, on average, by a factor 2. 
Since the typical fitness of a local optimum is such that
$1-F=O(1/N)$, and since the typical fitness attained after $\ell$
successful steps is of order $2^{-\ell-1}$, it follows that the typical
number of mutations after which an optimum is attained goes as
\be 
\ell_{\rm typ}\approx {\log N\over\log 2}. \ee
To translate this into time, we have to take into account that the probability
that the next move is successful is halved on the average at every time
step. It is then possible to show that the probability $Q_t$ that the
walk lasts $t$ generations is exponential:
\be 
Q_t\approx {1\over \ov{t}}\exp\l(-t/\ov{t}\r), \ee
where $\ov{t}=N$.

In the other extreme case of the Fujiyama landscape ($K=1$) one
obtaines instead
\be 
\ell_{\rm typ}\approx N, \qquad \ov{t}\approx N\log N. \ee
The proof is left as an exercise.

We now turn to considering the fluctuations in the reproductive
process \cite{WH}. Again the strong selection limit
is considered, thus, if a favorable mutation appears, it spreads
instantaneously into the whole population, but the fluctuations in the
reproduction are taken into account. This implies a finite probability
of {\em stochastic escape\/}
from the peak in fitness landscape: $p\approx u^M$, where $M$ is the
number of individuals in the population and $u$ is the mutation rate
per individual and per generation. On the other hand, the probability
that at least one better genotype is found is given by $q\approx
1-a^M$, where $a=1-(1-F)u$ is the probability that the fitness of an
individual does not increase respect to the maximal fitness of the
population, $F$. Two evolutionary regimes are found:
\begin{description} 
\item[Adaptive walk:] For $q<p$ the dynamics is essentially driven
  by mutations that increase the fitness, until a local fitness optimum is
  found.
\item[Stasis:] For $q\approx p$ the
  adaptive dynamics is very slow, and the genomic changes in the population
  take place because of random mutations that do not increase the
  fitness. Through this mechanism, the
  local optimum in the fitness landscape is eventually
  left by stochastic escape and a new adaptive phase begins.
\end{description}

In this simple model, the evolution shows the features of {\em
  punctuated equilibrium}.
The name {\em punctuated equilibrium\/} was proposed by S.J. Gould and
  N. Eldredge \cite{EG} to describe a characteristic feature of the evolution
  of simple traits observed in the fossil record. In contrast with the
  gradualistic view of evolutionary changes, these traits typically
  show long periods of stasis interrupted by very rapid changes.

One can interpret this phenomenon in two ways:
\begin{enumerate}
\item In microevolutionary models, punctuated equilibria can be thought of
  as a consequence of the complex dynamics of evolution. The periods
  of stasis are interpreted as metastable states of the population,
  and the rapid changes as barrier crossings, where the barriers are
  either ``energetic", as in the present model (a local optimum of the
  fitness has to be left) or ``entropic", related to the fact that
  some traits are represented in an overwhelming portion of the genome
  space. This latter case applies in some models of RNA evolution, where it is
  possible to investigate the relation between ``genotype" (the RNA
    sequence) and phenotype (the RNA three-dimensional structure)
    \cite{Schu,Schu2}.
\item In macroevolutionary models a period of stasis in the
  evolution of a species can still be thought of as a metastable
  state of the dynamics of that single species. If the
  ecosystem is at ``equilibrium'', all its species are stable.
  However, if one species undergoes a change, it will also change
  the fitness landscape of the interacting species, generally
  leading to their destabilization. Thus an ``avalanche '' of
  evolutionary change will sweep through the ecosystem.
 \end{enumerate}

If we plot the fitness of the population as a function of
time, we find very rapid adaptive walks to a high fitness value, followed by
long periods of stasis, which eventually end either by the discovery
of a better genome or by stochastic escape. In the last case the
fitness decreases and the adaptive process restarts. It is interesting to note
that an increase of the size of the population has opposite effects on
the rate of evolution in the two phases: the rate is increased in the
adaptive phase and decreased during the stasis periods.

This model is very simple but rather complete, since it takes into
account many of the basic ingredients of the microevolutionary models:
the ruggedness of the fitness landscape, the effects of mutations and
the finite size of the population. Despite its simplicity, it is able
to capture some generic features of the evolutionary process.

\section{Coevolution}
We have considered so far evolution taking place in a {\em fixed\/}
fitness landscape. Even in the case of a single population, this is a
drastic oversimplification. Genetists considered long ago models
where the fitness depends on the state of the population (frequency
dependent fitness). Models with a fixed fitness landscape describe a
situation where there is no interaction between individuals (apart for
the constraints due to limited resources) and are unable to describe
a situation where more than one species is present. We now consider
the modeling of interactions between species (coevolution). If one considers two
interacting species, one may have
three possible situations~\cite{MS}:
\begin{description}
\item[Competition:] the presence of each species inhibits the
  population growth of the other.
\item[Exploitation:] The presence of species A stimulates the
  growth of species B, and the presence of species B inhibits the
  growth of species A.
\item[Mutualism:] the presence of each species stimulates the
  growth of the other.
\end{description}
The host-parasite and the prey-predator interaction are well known
cases of exploitation. Exploitation leads, in a physical language,
to {\em frustrated\/} systems, in which it is difficult to reach
stable equilibrium. Prey-predator interactions lie at the origin
of quantitative population theory via the classic work of Lotka
and Volterra \cite{Lotka,Volterra}, which lies however outside
of the scope of these lectures. The essential lesson one draws from
their approach is that one expects situations where the size of the
interacting populations varies cyclically. One can ask what implications
appear in the genotype distribution.
One can build up a simple model \cite{Clarke} by considering one gene for the
host (or prey), one for the parasite (predator), and two
possible {\em alleles\/} (alternative forms) for each.
Let us denote these alleles by $(+,-) 1$.
One assumes that the combinations $({+}{+})$ and $({-}{-})$
are favorable to the parasite and unfavorable to the host, say
by multiplying the parasite's fitness by $(1+s)$ and the host's
one by $(1+r)^{-1}$. We can denote the frequency of the allele $+$
in the host population by $H_+$ and the corresponding quantity for
the parasite by $P_+$. In a large population,
the probability of an encounter of a $+$-host
with a $+$-parasite is proportional to $H_+P_+$.
This leads to a number of host offspring
proportional to $H_+P_+/(1+r)$ and to parasite
offspring proportional to $(1+s)H_+P_+$.
Neglecting the effects
of mutations, one thus obtains an equation of the form
\begin{eqnarray}
H_+(t+1)&=&H_+\frac{1+r(1-P_+(t))}{1+r(H_+(t)+P_+(t)-2H_+(t)P_+(t))};\\
P_+(t+1)&=&P_+\frac{1+sH_+(t)}{1+s(1-P_+(t)-H_+(t)+2H_+(t)P_+(t))}.
\end{eqnarray}
It is easy to see that this equation leads to oscillating behavior,
akin to that of the Lotka-Volterra equations. If one introduces
mutations, oscillatory behavior only sets in if $r$ and $s$ are large enough.

This picture is reminiscent of
the mechanism postulated by Van Valen \cite{VV} to explain the fact
that the number of species surviving longer than a time $t$
decays exponentially with $t$, as if the probability of extinction
were independent of the age of the species (and thus of the degree of
adaptation reached). Van Valen called this mechanism the Red Queen
Hypothesis, after the episode of Lewis Carroll's book ``Through the
looking glass" where the Red Queen explains to Alice that she has to
run very fast if she wants to remain in the same place. In the
evolutionary language the metaphor means that the evolutionary changes
are mainly ``aimed" to avoid to get extinct in an ever
deteriorating environment, rather than to improve the fitness in a
stable environment (or fixed fitness landscape, in the language
of population genetics).

A model of the host-parasite interaction was proposed by Hamilton et al.
\cite{Ha} in order to investigate the advantages of sexual reproduction
over the asexual one. Despite its success in nature, sexual
reproduction is not equally successful in mathematical models. One of
the main disadvantages that appear in the models is the ``cost of
males": since in an asexual population all the individuals produce offsprings,
whereas in sexual reproduction only half of the individuals are able to
do that, an asexual population should in principle be able to
reproduce much faster than a sexual one. This fact must be compensated
by some advantage for sexual reproduction. But, in models with fixed fitness
landscapes, the sexual reproduction is favored only in very special
situations. On the contrary, in an environment that is rapidly changing
because of parasites that keeps on mutating, favorable mutations
spread much faster through sexual reproduction than through the
asexual one. This explanation of the advantage of sex is inspired by the
  same kind of philosophy as the Red Queen hypothesis.
Through sexual reproduction, favorable mutations taking
place in different individuals are rapidly assembled in an unique
genome. Moreover, the fixation of a favorable mutation in an asexual
population requires that all the less fit individuals eventually die without
leaving offspring, while in a sexual population it is enough that all
of the females of the population couple with a male that bears the mutation.

Through numerical simulations of his (rather complex!) model, Hamilton and
collaborators were able to show that the mechanism that he proposed gives to sexual
reproduction an advantage large enough to compensate for the cost of
the males. However, several other mechanisms have been proposed in the
literature to explain the ubiquity of sexual reproduction, and the
problem is far from being settled.

Of course, in nature, interactions among species are not restricted to
isolated pairs. One should instead imagine each species as a member
of an interaction web, and the treatment of the problem becomes
rather quickly extremely involved. One can hope to simplify it
by carrying over to the coevolutionary case some of the approaches
that we have discussed for the evolution of a single species.
For example, Kauffman \cite{Ka} has introduced a model,
called the NKC model, which is a generalization of his NK model of adaptive
walks. There are $S$ interacting species, each of which is represented
by a point in its own genome space. The genome length is $N$ for every
species. The fitness of a genome ${\bf s}^\a$ in species $\a$ depends on
${\bf s}^\a$ itself and on the state of $C$ randomly chosen elements in the genomes
of other species:
\be 
\hbox{Fitness}({\bf s}^\a)=F_K^{(\a)}\l({\bf s}^\a,s_{i_1}^{(\a_1)}, \ldots
s_{i_C}^{(\a_C)}\r). \ee
As usual, $K$ measures the ruggedness of the landscape, going from
$K=1$ (each genomes elements contributes independently of the others)
to $K=N-1$ (Random Energy Model).

Each of the $S$ species performs an adaptive walk in its own genome
space, where the fitness landscape depends on the state of
the other species. After a transient time the fitnesses of all species
reaches a metastable state where the mutation of the species would
lower its own fitness. This is a very fragile equilibrium, since there
is no global function being optimized: every species reach a point
which is a local optimum {\em provided that the other species do not
  mutate}. This kind of state is known in the economic theory as
Nash equilibrium.

The Nash equilibrium is reached after a very long time even with a
small number of species, the longer the smaller is $K$. The fitness
reached is very low and tends to the average value as $K$
increases. The interactions between species are frustrated, like in
the host-parasite problem, but it is also possible to observe
cooperative effects.

A solvable version of this model was proposed by Bak, Flyvbjerg
and Lautrup \cite{BFL}. In this model, the fitness
landscape is completely rugged and
the $C$ interacting species are chosen anew at each time step. In the
language of disordered systems, one has an {\em annealed\/}
approximation to Kauffman's model. One obtains analytical
results for $N\gg 1$ and for the number of species $S\to\infty$.

Let us denote by $\rho_M(F,t)$ the fraction of species at time $t$ with
fitness $F$ and in a position such that $M$ mutations decrease their
fitness. The probability that a mutation is accepted is then
\be A(t)=\sum_{M=0}^N \l(1-{M\over N}\r)\int_0^1 dF \rho_M(F,t). \ee
This quantity is called the {\em activity\/} of the system, and  
discriminates the possible behaviors.
It is possible to derive a master equation for $\rho_M(F,t)$.
Let us define two more
quantities:
\begin{itemize}
\item $\Phi(F,t)$ is the probability that a mutation is accepted, and
  results in a new fitness equal to $F$. One has
\be \Phi(F,t)=\sum_{M=0}^N \l(1-{M\over N}\r)\int_0^F dF'
{\rho_M(F',t)\over 1-F'}. \ee
\item $B_{M,N}(F)$ is the probability that $M$ possible mutations of a
  genome with fitness $F$ have a fitness lower than $F$. This is given
  by $B_{M,N}(F)={N\choose M} F^M(1-F)^{N-M}$.
\end{itemize}
With these notations, one obtains the following master equation:
\begin{eqnarray}
 {\pa\over \pa t}\rho_M(F,t)&=&-\l(1-{M\over N}\r)
\rho_M(F,t)+B_{M,N}(F) \Phi(F,t)-{c\over N} A(t)\rho_M(F,t)\nonumber\\
&&+{c\over
  N}A(t) B_{M,N}(F) \label{master}. \end{eqnarray}
This equation exhibits two behaviors, depending on the connectivity $C$ of the
ecosystem:
\begin{enumerate}
\item For $C< C_{\hbox{crit}}$ one has Nash equilibrium. The condition of the
  Nash equilibrium reads
\be \rho_M(F,t)=\d_{M,N} \rho(F). \ee
In other words, no mutation should lead to the improvement of the fitness
of any of the existing species.
\item For $C>C_{\hbox{crit}}$ one has a ``Red Queen" phase . In this case, 
there is a  non-trivial stationary solution, with an activity $A^*$ different
  from zero. In other words, no stable Nash equilibrium can be
  reached and all the species keep constantly mutating.
\end{enumerate}

The instability of the Nash equilibrium in the ``Red Queen" phase
can be understood through the following argument: the number of genes
changed on the average in an adaptive walk of an isolated species is
given by
\be \mu_1=\log N+{\rm const.}+ O(1/N). \ee
If this change forces the evolution, on average, of more than one species, a
chain reaction starts that makes the Nash equilibrium
impossible. Since on the average $C$ species receive inputs from a
given species, and the probability that the input comes from a
mutating element is $\mu_1/N$, the critical connectivity is given by
\be C_{\hbox{crit}}=N/\mu_1\simeq N/\log N. \ee

This means that, if more than a fraction of the genome of order
$1/\log N$ is influenced by the other species, no stable Nash
equilibrium can exist (of course this conclusion is strongly dependent
on the assumption of a completely rugged fitness landscape). The
activity of the system at stationarity, that is the order parameter for this
transition, can be computed self-consistently \cite{BFL}.

This description holds at a coarse-grained level at which the
population is represented as a single genome. The effects of the
variability inside the population have not been explored. It can be
conjectured that, in the framework of the quasi-species theory, a
transition from the equilibrium to the Red Queen phase would be
observed, analogous to a statistical mechanics transition. If a
finite, asexually reproducing population is represented, we would
expect the Muller ratchet mechanism to destabilize the Nash
equilibrium also in the small $C$ regime where it is stable in the
above approach.

A final remark about the role of the fitness in the above model is
due. Fitness was introduced in section \ref{Darw} as a quantity
proportional to the {\it relative\/} reproductive rate of individuals sharing a
given genome in a given population. The average fitness of a
population $\a$ in an ecosystem of many populations is {\em not\/} the
analogous of the fitness at the individual level, in the sense that it
has nothing to do with the probability that a given species thrives at
the expense of other species.
The role played by the fitness in the dynamics of this coevolutionary
model is thus different from the one played in
microevolutionary models, and two different situations have to be
distinguished: if none of the $C$ species with which a given species $\a$
interacts is mutating, then the probability that a mutation is
accepted decreases as the fitness of the species $\a$ increases. This
situation takes place near a Nash equilibrium. If, on
the other hand, one of the species that constitute the environment of
the species $\a$ changes, then the following evolution is completely
independent of the previous value of $F_\a$ (this holds in the
framework of the present model, that assumes a completely rugged
fitness landscape).

The models of macroevolution, which we will discuss in the next
section, can also be distinguished according to which situation is
assumed. The Bak-Sneppen model \cite{BS} assumes that the system is near to a
Nash equilibrium (the connectivity of the model, $C$, is small), so
that the time-scales for mutations are very different from one population
to another (and are related, in this interpretation, to the fitness of
the population). Thus one assumes that, at every step, only the population
with the smallest time-scale is allowed to mutate. This mutation
may destabilize the populations connected to the one that mutates, and
propagate in the system as an {\it avalanche}. Other models assume a
much faster varying environment, so that the notion of Nash
equilibrium is much less relevant.

\section{Macroevolutionary patterns}
We now consider large-scale evolution. The evolving units will
now be typically species, and we assume that, in the
large-scale records the relevant facts are the presence or absence of a
 species, and not its size (although
the number of individuals in a group might have a role in survival). 
I have pointed out before that the biological concept of {\em
fitness\/} cannot be straightforwardly
applied to whole species. In the
frame of macroevolution, the survival probability of a species does not
simply depend on
the reproduction rate of its individuals. In fact there is much debate on the
correct level of description needed when one
considers large-scale evolution. We shall first recall
a few simple facts about evolution in the biosphere,
and we shall then describe different approaches that
aim at explaining the apparent patterns of macroevolution
in terms of simple, robust, underlying mechanisms. 

Let us start then with data for extinction and evolution of
pluricellular life on Earth. Our record starts 600 My ago, near what
is termed ``the Cambrian Explosion". Before that time, life had been
represented by simple unicellular organisms. But with
pluricellular life, different more complex organisms started to
develop. In a few million years, many different corporal plans were
explored, including different symmetries (triradiated bodies, for
instance, or soft-bodied marine animals with five eyes) and
architectures that cannot be found nowadays (see the excellent book by
Stephen Jay Gould ``Wonderful life" for an insight into this lost
world \cite{Gould}. Since then,
diversity (usually computed as the total number of taxonomical
families) has been always increasing on the average, although
interrupted by a few large mass extinctions. The organisms on Earth can be grouped
taxonomically, that is, a certain number of closely related species form
a {\em genus}, and genera are grouped into {\em families}. These are the three
taxonomical levels relevant for the following discussion. When discussing
the data, it is worthwhile to keep in mind 
that the further in the past we look, the less reliable our data
become, and that it is always easier to get good data for a family that
for a genus or a species. For instance, it is easier to spot the
extintion epoch of a genus than of a species: just one species is enough to
establish its presence, while all the species in it should have died
out in order to say that the genus has gone extinct. It would be 
even better to compute the same quantity for families: nevertheless, statistical
problems may arise due to the fact that when going one level higher,
the number of data decreases by roughly an order of magnitude. 

The following observations, made in the last years, could serve as a
starting point in the formulation of simple models of macroevolution: 
\begin{enumerate}
\item The distribution $N(m)$ of extinction sizes for families $m$
 decreases with $m$ according to a power-law: $N(m) \propto m^{-\alpha}$ with $\alpha
  \simeq 2$ (see, e.g. the book by Raup \cite{Raup}).
   Although often quoted as the paradigm of
  critical behavior in macroevolution, this data set has probably the
  largest error of all the observations \cite{PRS}.
\item The distribution $N(t)$ of genera lifetimes $t$ follows a 
  well defined power-law $N(t) \propto t^{-\kappa}$ with 
  $\kappa = 2.10 \pm 0.11$ \cite{Sep,Sep2}.
\item The statistical structure of taxonomy follows clearly defined
  laws that do not depend on the level at which we are looking at or
  on the particular case group. For instance, the distribution of
  the number of genera $N(g)$ with $g$ species, or number of families
  with $g$ genera are the same and can be fitted (again) with
  a power-law with exponent close to -2 \cite{Burlando}.
\item The probability for a group (in the taxonomical sense) to
  disappear in a given time interval does not depend on its lifetime: 
  the rate at which species or genera 
  disappear reminds us of radioactive decay, in
  which the amount of the ``original element" decays exponentially
  with time. This result, first reported by van Valen \cite{VV},
  and related by him to the Red Queen effect,
  strongly corroborates the idea
  that, even if the fitness of individuals might increase through
  evolution, it not correlated with the survival probability of a
  species.
\item The punctuated pattern of extinction events is not random but
  displays long-time correlations. The study of different
  paleontological measures along 600 My reveals the presence of $1/f$
  noise and non-trivial correlations for families extinction, and 
  diversity fluctuations in genera and species of particular 
  groups \cite{Nat}.
\end{enumerate}

According to these observations, we remark that the so termed ``big
five" extinctions appear to belong to the tail of a very skewed 
 distribution, as stressed in 1.\ above. This point requires
however further study. We should keep in mind that
 we are dealing with a highly non-stationary system:
for instance, it appears that the extinction rate is decreasing through
time, while diversity is increasing. This is not in
  contradiction with the Red Queen effect, according to which the
  probability of extinction for a certain group $i$ does not depend on its
  age, that is $N_{t+1}^i=\alpha N_t^i$, with $\alpha < 0$. In fact, if
  the rate of appearance of new groups {\it does\/}
  depend on time (if, say, $M_t=\sum_i
  N_t$ grows as $M_{t+1}=\beta (t) M_t$, where $\langle \beta (t)
  \rangle > 1$) then the global extinction
  rate $\gamma$ might decrease, $\gamma=\alpha / \left((1-\alpha) +
      \beta (t)\right)$, while remaining constant over time within
      each group.
Moreover, due to the obvious difficulty of sampling, our data might be
incomplete in a significant way, although it seems that statistics 
have not been strongly changed by 20 years of intensive data 
recruitment, see \cite{Mike}. 

An alternative way of looking at some data is offered by the {\it
killing curve}. It is defined as the integrated number of killed
species (in percent) vs.\ the mean waiting time between
extinctions. We get a sigmoidal curve with the largest extinctions at 
the top. See for instance \cite{Raup} and, for
its potential applications to modeling \cite{New1}. 

The overwhelming majority of data correspond to hard-bodied organisms,
meaning that data are often lacking for plants, for instance, which are
only seldom preserved. Some old data by Yule \cite{Yule} yield
an exponent close to $3/2$ for the taxonomical hierarchy in plants,
but further information is lacking.

\section{Models of macroevolution}
The presence of scaling laws in macroevolutionary data led to the
supposition that the dynamics of large-scale evolution could be the
result of a self-organized critical process. In 1993, Bak and Sneppen
introduced a toy model (BS) for ``species" evolution \cite{BS}. In its
original version, $N$ species are arranged in a one-dimensional lattice,
and a real number $x_i$ between 0 and 1 is assigned to each. 
The value of $x_i$ represents the height of the barrier that species $i$,
($i=1,\ldots,N$) must overcome in order to
mutate.If we assume that
he time to overcome a barrier depends exponentially on the
barrier height, it follows that only the species with the
lowest barrier, let us say $x_{\rm min}$, is able to mutate. 
The model does not give an interpretation of species mutation:
it can be real extinction, where the sepecies is replaced by the
descendants of a different species, or a pseudoextinction,
where the species is replaced by its own descendants, but with
rather different characters. Because of the species mutation,
the environment of the species and of the ones with which it
interacts changes. In order to model this effect,
a new random number is given
to the new species and to its two nearest neighbors. As a consequence,
one of the three species whose $x$ has been recently
drawn is more likely to mutate at the next step.
If this is the case, one has an ``evolutionary avalanche''.
When, by chance, the last evolved species have a barrier
so large that $x_{\rm min}$ belongs to none of them, the
avalanche stops.

The BS model self-organizes close to a critical point at which the
evolutionary activity is barely maintained. Almost all species have
barriers above a critical threshold $x_{\rm c} \simeq 0.7$,
and $x$ lies below the threshold for
just one species on average. One
obtains self-similar distributions for relevant quantities, and some
of the features of macroevolution are qualitatively recovered: there is
punctuated equilibrium behavior and the distribution of avalanche
sizes is described by a power-law with an exponent $\alpha_{BS} \simeq 1.1$, 
which is however quite far from the observed value. 
The model has an annealed version that
can be analytically solved \cite{BSF,deB}. There, the
neighbors of the cell with the minimum barrier are chosen at random
and the model can be mapped onto a branching process: the resulting
exponent for the distribution avalanche sizes is 3/2. 

Since this first model, many others have been proposed. Some of them
do not take self-organized criticality as a requirement for scaling
and others do not put internal dynamics as the main force to shape
large-scale patterns. Mark Newman \cite{New} has considered the
role played by external causes and has presented a simple model of
non-interacting species with results compatible with the observed distribution
of extinction sizes. In his approach, he considers $N$ species
characterized, like in the BS model, by a real number $x_i \in [0,1]$,
$i=1, \ldots, N$. The external {\it stress\/} $s(t)$ is a random variable 
drawn
from a certain decreasing distribution (an exponential, a power-law,\dots\,),
with the single requirement that its average is closer to 0 than
to 1. At each time step, all the species such that $x_i< s(t)$ are
removed and replaced by new ones, with a random number chosen from a uniform
distribution. There is also some small internal change to prevent the
system from freezing: a small fraction of the species is also changed at
random at each time step. 

Although the emphasis in Newman's model is put on external causes,
he has quite remarkably found that the 
quantitative distribution of extinctions  
(computed simply as the number of species removed at each time step) 
depends only weakly on the form of the external stress, and in
particular fits well the available data. The model, however, has
some shortcomings: in particular it does not involve taxonomy,
unless one decides to introduce it, quite arbitrarily,
by assigning to each newly introduced species a randomly
chosen ancestor species among the survivors.

In 1996, Sol\'e and Manrubia introduced an ecological
model of evolution and extinction in which the emphasis was put both
in internal dynamics and in the ecological web of interactions \cite{Sole,Sole1}. 
Each species is characterized by a set of inputs from the other
$N-1$, thus the relevant dynamical object is the matrix of connections
$J_{ij}$ formed by elements which take real values between -1 and 1. 
This matrix is updated as follows. First, one of the input connections
for each of the species is changed at random. Next, the sum of the
inputs to all the species is calculated, $h_i=\sum_j J_{ij}$, and if
$h_i$ falls below zero the species dies out. Its connections are
removed and are 
replaced by the links of a randomly chosen surviving species.
 This last step defines a natural taxonomy in the system. 
This model gives results compatible with field observations for the
distribution of extinction events, of lifetimes and of species within
genera. In a simple approximation \cite{MP}  one obtains analytically
the observed exponent $\alpha=2$
of the extinction size distribution. Moreover, it
gives macroevolution a different interpretation: in this model, the
Red Queen effect is also observed (species disappear at a rate
independent of their lifetime), but now a consequence of the race for life
and survival, but just  of the random mutations
that slowly weaken the ecosystem and push it to a
``critical state", in which small perturbations can trigger big
avalanches. 

The previous models obviously lack some ingredients that seem
essential to the real process, like the increase of the
total number of species or the fact that evolution appears
to be in a non-stationary
state. More recently new models have tried to overcome these drawbacks by
introducing, for instance, a variable system size \cite{Wilke,Head}, while some
others have considered the introduction of external
perturbations simultaneously to internal dynamics \cite{Sibani}.

One can notice that all the described models fall essentially into two
classes: one group considers the internal dynamics of
the system as the main cause of the observed regularities, while a second
group considers that external causes act not only as driving or 
triggering forces, but also as the real causes for the observed scaling
laws. In any case, the five observations listed in the previous
section should be consistently recovered by any reliable model of
macroevolution. There are probably also some relations among them that
a realistic formulation should be able to identify
and explain. And still the debate of
the ``modelizability" of macroevolution is open, and some authors
(mainly paleontologists) see so many different causes and so many
interacting variables, that they will never concede that a simple model will
be able to account for the gorgeous variability of the history of
life.

\section*{Acknowledgments}
The backbone of these lectures arose from
correspondence with Paul G. Higgs, whom I thank for 
having shared with me many of his
ideas. The responsibility of any distortion and misinterpretation lies
on me alone. I also thank the Laboratoire de Physico-Chimie
Th\'eorique, ESPCI, Paris, for hospitality and encouragement
during the preparation of these lectures.
I am grateful to U. Bastolla and S.C. Manrubia for the great
work they have done in taking and writing down these notes.

\end{document}